\def\year{2022}\relax
\documentclass[letterpaper]{article} 
\usepackage{aaai22}  
\usepackage{times}  
\usepackage{helvet}  
\usepackage{courier}  
\usepackage[hyphens]{url}  
\usepackage{graphicx} 
\usepackage{natbib}  
\usepackage{caption} 
\DeclareCaptionStyle{ruled}{labelfont=normalfont,labelsep=colon,strut=off} 
\usepackage{algorithm}
\usepackage{algorithmic}

\usepackage{amsmath}
 
\usepackage{newfloat}
\usepackage{listings}
\floatstyle{ruled}
\newfloat{listing}{tb}{lst}{}
\floatname{listing}{Listing}
\title{Predicting infections in the Covid-19 pandemic - lessons learned}
\author{
  Sharare Zehtabian, 
  Siavash Khodadadeh, 
  Damla Turgut,
  Ladislau B{\"o}l{\"o}ni
}
\affiliations{
    Department of Computer Science, University of Central Florida\\
    Orlando, Florida, 32816\\
    \{sharare.zehtabian, siavash.khodadadeh, damla.turgut, ladislau.boloni\}@ucf.edu
}

\begin{document}

\maketitle

\begin{abstract}

Throughout the Covid-19 pandemic, a significant amount of effort had been put into developing techniques that predict the number of infections under various assumptions about the public policy and non-pharmaceutical interventions. While both the available data and the sophistication of the AI models and available computing power exceed what was available in previous years, the overall success of prediction approaches was very limited. In this paper, we start from prediction algorithms proposed for XPrize Pandemic Response Challenge and consider several directions that might allow their improvement. Then, we investigate their performance over medium-term predictions extending over several months. We find that augmenting the algorithms with additional information about the culture of the modeled region, incorporating traditional compartmental models and up-to-date deep learning architectures can improve the performance for short term predictions, the accuracy of medium-term predictions is still very low and a significant amount of future research is needed to make such models a reliable component of a public policy toolbox.

\end{abstract}


%

\section{Introduction}

There is no epidemiological event in the history of humanity that received the level of computational modeling and prediction effort as the Covid-19 pandemic. In contrast to previous outbreaks, humanity faced the pandemic not only with the scientific models honed over many previous outbreaks but also with new computational tools and a relatively well functioning international reporting system. The latter, at least in the developed world, allowed the accurate tracking of the infections, hospitalizations, and deaths on a day-by-day, region-by-region basis. Gene sequencing tools allowed the tracking of virus variants in each area. Finally, the pandemic outbreak occurred after the deep learning revolution in AI, opening the possibility to learn predictive models from the data as it becomes available.


Yet, with all these apparent advantages, humanity's ability to predict the spread of infections was little better than random. Repeatedly, public policy measures were based on wildly incorrect forecasts. Predicted waves did not materialize, while significant outbreaks happened in locations and at times that nobody predicted. As different variants of the virus had different levels of infectiousness, authorities were effectively dealing with several overlapping pandemics, connected by the partial immunity conferred by previous infections. The change in recommended treatment procedures, the emergence of vaccines, and more recently, antiviral medications also changed the nature of the pandemic. The strong age-dependency of the medical outcomes meant that the pandemic waves were dependent on the age pyramid of the geographic area. Finally, countries and regions instituted different non-pharmaceutical interventions (NPIs): public health measures, lockdowns, masking recommendations, and so on. The nature and stringency of these measures depended not only on the evolution of the pandemic but also on local political events, popular opinion, and personalities with strong influence over the public narrative. 

Despite the weak results of the pandemic prediction approaches, continued research into the prediction models remains highly important for the remainder of the Covid-19 pandemic, as well as for future epidemics. Even just a better understanding of what is predictable and what is not would be helpful for policymakers of the future. 

The impetus for this paper started from the Pandemic Response Challenge sponsored by Cognizant Inc. at the XPrize Foundation, launched on October 30, 2020. The two phases of the competition focused on prediction and prescription models respectively. For the prediction phase, the competition organizers implemented a framework in which the prediction models submitted by the contestants have access to the Oxford Covid-19 Government Response Tracker (OxCGRT) data~\cite{hale2021global}. Several implementations of possible predictor architectures, such as an LSTM-based model proposed in~\cite{miikkulainen2021prediction} were included in the starting package. We will refer to this model, developed at the University of Texas Austin and Cognizant Inc. as the LTSM-UT-Cogn model in the rest of this paper. For the first phase of the competition, a leaderboard was dynamically evaluated every day on up-to-date real-world data from Dec 22, 2020, to January 12, 2021. The final phase of the competition focused on choosing an optimal set of NPIs, which, while interesting from an algorithmic point of view, cannot be evaluated with real-world data.

The authors of this paper entered the challenge as the PawP (Pandemic Wave Predictor) team with an earlier version of the model referred to as LSTM-CultD-SIR in this paper. Our highest ranking in the leaderboard was on January 14, 2021, where our team was 5th out of 336 registered teams. Since the closing of the competition, the pandemic went through several new phases. For instance, in the United States, the availability of vaccines lowered the number of cases in June 2021 to a point where the public perception considered the pandemic to be over. Then, the spread of the Delta variant lead to a new wave of infections in Fall 2021. However, the availability of treatments such as monoclonal antibodies lowered the mortality rate compared to previous phases of the pandemic. On the other hand, ``pandemic fatigue'', economic pressures and cultural backlash affected the ways NPIs are chosen as well as the population compliance level. 

The objective of this paper is to revisit, 11 months later, prediction models that were competitively among some of the most accurate ones in January 2021, and investigate whether any useful lessons can be learned from the evolution of the pandemic since then. 
 The overall questions we want to answer in this paper are the following:

\begin{itemize}
    \item How do prediction models that were successful in the early stage of the pandemic hold up at later stages? Clearly, the prediction models trained on data from the early part of the pandemic will be useless in later phases. It is reasonable to ask, however, whether the learning architectures, trained with new data, would retain their respective advantages?
    \item The public press assigned a significant importance to cultural factors for the different evolution of the pandemic in specific countries or regions. Can we gain additional accuracy (in the long run) by using quantifiable descriptors of culture in various countries and regions as inputs to our model?
    \item Would more advanced machine learning techniques improve the accuracy? For instance, would the most recent advances in AI models, such as the multi-head attention model used in transformer models~\cite{vaswani2017attention} help the accuracy of the prediction? 
\end{itemize}



\section{Related Work}
\label{relatedwork}


The Covid-19 pandemic led to the initiation of a significant number of modeling and prediction projects in the academic community. 


\cite{flaxman2020estimating} studied the effects of major non pharmaceutical interventions (NPIs) across 11 European countries and introduced a Bayesian model to estimate the epidemic. Their model calculates backwards from observed deaths to estimate transmission that occurred several weeks previously, allowing for the time lag between infection and death.


\cite{dehning2020inferring} focused on short-term infection forecasting based on NPIs and studied how the interventions affect the epidemiological parameters. They combined a SIR model with Bayesian parameter inference to analyse the time-dependent spreading rate. They detected the change points in the spreading rate that have correlations with the times of announced interventions. They specifically focused on Covid-19 spread in Germany. 

\cite{arik2020interpretable} proposed an approach for modeling Covid-19 forecasts by integrating covariate encoding to compartmental models. They used the standard SEIR model but modeled more compartments such as: undocumented infected and recovered compartments, hospitalized, ICU and ventilator compartments, and partial immunity.


\cite{jin2021inter} focused on a direct data-driven prediction model for predicting Covid-19 without using compartmental models. They developed a neural forecasting model called Attention Crossing Time Series (ACTS) that predicts cases by comparing patterns of time series from different regions. They addressed the scarcity of time series historical data for each region by investigating other regions’ time series in the dataset with similar long term or short term patterns. 

\cite{xiao2020c} proposed a data-driven framework called C-Watcher to screen all the neighborhoods in a city and detect the neighborhoods with the highest risk of Covid-19 spread before they contaminate other neighborhoods. They used long-term human mobility data from Baidu Maps and characterized each neighborhood by using urban mobility patterns. 

\cite{liao2020tw} proposed a time-window based SIR prediction model and used a machine learning method to predict the basic reproduction number $R_0$ and the exponential growth rate of the epidemic. For their time-window based sir model, they specifically split historical data into a time window segment in order to capture the real-time changes in $R_0$ and the exponential growth rate. They used Covid-19 historical data in China, South Korea, Italy, Spain, Brazil, Germany and France. 

\cite{mehta2020early} focused on country level prediction of Covid-19 for the near future based on a combination of health statistics, demographics, and geographical features of counties. They used US Census data to obtain county-level population statistics for age, gender, and density.
The project used an XGBoost classifier to classify each county either as a positive or negative instance. To predict the number of occurrences, they used a XGBoost regression model. Finally, they combined results from the first two stages and calculated the expected occurrences for counties as a measure of county vulnerability.

\cite{watson2021pandemic} proposed a Bayesian time series model that fits a location-specific curve to the velocity (first derivative) of the log transformed cumulative case count. Then, they use a random forest algorithm that learns the relationship between Covid-19 cases and population characteristics to predict deaths. Finally, they embed these models to a compartmental model which can provide projections for active cases and confirmed recoveries. 

\cite{zou2020epidemic} introduced the SuEIR model, a variant of the SEIR model, to predict confirmed and fatality cases, the peak date of active cases, and estimate the basic
reproduction number ($R_0$) in the United States. Their model considers additional information such as the untested/unreported cases of Covid-19 and is trained by machine learning based algorithms by using historical data. They fit an ordinary differential equation (ODE) based model on the data. Their model could provide accurate short-term (daily ahead) projections for both confirmed cases and fatality cases at national and state levels. In the long term, they showed that the numbers of confirmed cases and deaths will keep increasing rapidly within one month.

\cite{qian2020and} focused on developing a model to learn the policies that affect the fatality rate of the Covid-19 in a global context. They used a Bayesian model with a two-layer Gaussian process (GP) prior. The lower layer models the Covid-19 fatality curve over time within each country with a compartmental SEIR (Susceptible, Exposed, Infectious, Recovered) model. The upper layer is shared with all countries and it is another GP model that learns the $R_0$ as a function of country features and policy indicators.  

\cite{sharma2020robust} investigated the robustness of the estimated effects of NPIs against Covid-19. In particular, they studied if NPI effectiveness estimates generalize to unseen countries without access to the ground truth NPI effectiveness estimates. 

\cite{mastakouri2020causal} studied a causal time series analysis of the Covid-19 spread in Germany to understand the causal role of the applied non pharmaceutical interventions (NPIs) in containing the spread among German regions. They used a causal feature selection method for time series with latent confounders called SyPI to analyse and detect the restriction policies that have a causal impact on the daily number of Covid-19 cases. They performed the analysis on a state and on a district level. 

\cite{yeung2021machine} combined NPIs and Hofstede cultural dimensions in predicting the infection rate for 114 countries. Particularly, they predict confirmed infection growth (CIG), which is defined as the 14-day growth in the cumulative number of reported infection cases. They used OxCGRT data of the NPIs and trained different non-time series models such as ridge regression, decision tree, random forest, AdaBoost, and support vector regression using mean squared error (MSE), and performed a grid search on the combination of these models. 

\cite{johndrow2020estimating} built a model for Covid-19 transmission only by using the number of daily deaths, timing of containment measures, and information on the clinical progression of the disease. The authors suggested that using less precise information such as the number of confirmed cases will result in unreliable analysis.  Their modeling approach is a SIR model of disease spread via a likelihood that accounts for the lag in time from infection to death and the infection fatality rate. They specifically fit a Markov Chain Monte Carlo (MCMC) algorithm to their data.

\cite{bengio2020predicting} proposed a proactive contact tracing method. They embedded two neural networks namely Deep Sets and Set Transformers and evaluated the resulting models via the COVIsim testbed. Their methods are able to leverage weak signals and patterns in noisy, heterogeneous data to better estimate infectiousness compared to binary contact tracing and rule-based methods. Their method provides a good trade-off between restrictions on mobility and reducing the spread of disease.

\cite{davies2020effects} designed a stochastic age-structured transmission model to study different intervention scenarios and their impacts on the transmission of Covid-19 cases in the UK. They specifically explored four base interventions scenarios of school closures, physical distancing, shielding of people aged 70 years or older, self-isolation of symptomatic cases, and modelled the combination of these interventions.

\section{Learning-based models for predicting the number of infections}
\label{pandemicpred}

\newcommand{\npi}{{\mathit{NPI}}}
\newcommand{\npimax}{{\mathit{NPI\_MAX}}}
\newcommand{\predictor}{\mathcal{P}}
\newcommand{\predictionratio}{\mathcal{r}}
\newcommand{\numcases}{n\mathcal{C}}
\newcommand{\numcasespredicted}{n\mathcal{\hat{C}}}
\newcommand{\prescripor}{\mathcal{Q}}
\newcommand{\costs}{{C}}
\newcommand{\population}{P}

In this section, we describe four alternative learning-based models for predicting the number of infected people in a given day of the pandemic in a particular area. The models assume that we have access to two streams of daily data. The {\em context stream} provides information about the total number of infected and dead people. This stream describes the basic context in which the prediction needs to be made - for instance, the total number of people already infected affects the number of people infected in the following days. We will include in this stream also the specific identifiers or the country and geographical region to which this data refers. The {\em action stream}, in contrast, describes the specific actions, typically non-pharmaceutical interventions that authorities enacted in a given area. Following the Pandemic Response Challenge setup, we extract the context and action stream from the Oxford Covid-19 Government Response Tracker (OxCGRT) data~\cite{hale2021global}. The action stream data includes 12 Non-pharmaceutical Intervention (NPI) columns: school closing, workplace closing, cancel public events, restrictions on gatherings, close public transport, stay at home requirements, restrictions on internal movement, international travel controls, public information campaigns, testing policy, contact tracing, and facial coverings. We also use context columns for each region, such as country name, region name, geo ID, date, confirmed cases, confirmed deaths, and population. We ignored countries or regions for which no number of cases or deaths are available and filled the empty or missing values on NPI columns in the data with 0 for each region or country.  

Finally, by having access to the country and geographical region, it is possible to extend the context and action stream with other information that can be looked up from other databases or web services. For instance, should we want to investigate the hypothesis that the weather affects the spread of an infection, this information could be, for example, brought in by correlating the geographical identifier with an external weather service. 

Having access to the stream of information contained in the context and the action stream and whatever auxiliary data the system might choose to look up, the objective of the predictive model is to predict the number of infections in the next day, and through extrapolation, for a larger period into the future. 

The notations used in this paper are as follows: for a region $r$ with population $\population^{r}$, we refer to the values of NPI columns for day $t$ by $\npi_{t}^{r}$ which is a vector of length 12. Each element $i$ is an integer between $0$ and $\npimax_{i}$. We denote the number of Covid-19 cases at day $t$ by $\numcases_{t}^{r}$.

\subsection{Learning based epidemiological models}



Traditional epidemiological models, such as the SIR compartmental model aims to predict the evolution of the epidemic based on first principles and a relatively few number of human-understandable parameters. The input to these models is usually the current state of the pandemic and they are calibrated by human experts based on the infectiousness of the disease. 

In contrast, learning based models use significantly more complex models with a very large number of parameters, such as deep neural networks. These parameters are not individually human-interpretable and the only realistic way to acquire them is through the use of learning. Often, these models look at the pandemic as a function unfolding in time, thus they take an input either a sliding window of the recent history of the pandemic at every timestep, or maintain internally a memory of it. 

The official examples and the majority of entries to the Pandemic Response Challenge were such learning-based systems (although it is difficult to know how much human-expert data was individually incorporated). In the reminder of this section we will discuss four possible models (a model presented as the ``official´´ one in the competition and three models designed by our team), presenting their architecture and design rationale in a comparable way (see Figure~\ref{fig:preds_methods}).

\subsubsection{LSTM-UT-Cogn}

The first model we are describing was developed by the organizers of the Pandemic Response Challenge and provided to the teams that qualified to the finals of the competition~\cite{miikkulainen2021prediction} to serve as a metric for prescriptive measures. Although this model did not directly compete in the challenge, it was clearly seen as a state-of-the-art model at that point in the pandemic. 

\begin{figure*}
    \centering
    \includegraphics[width=0.4\textwidth]{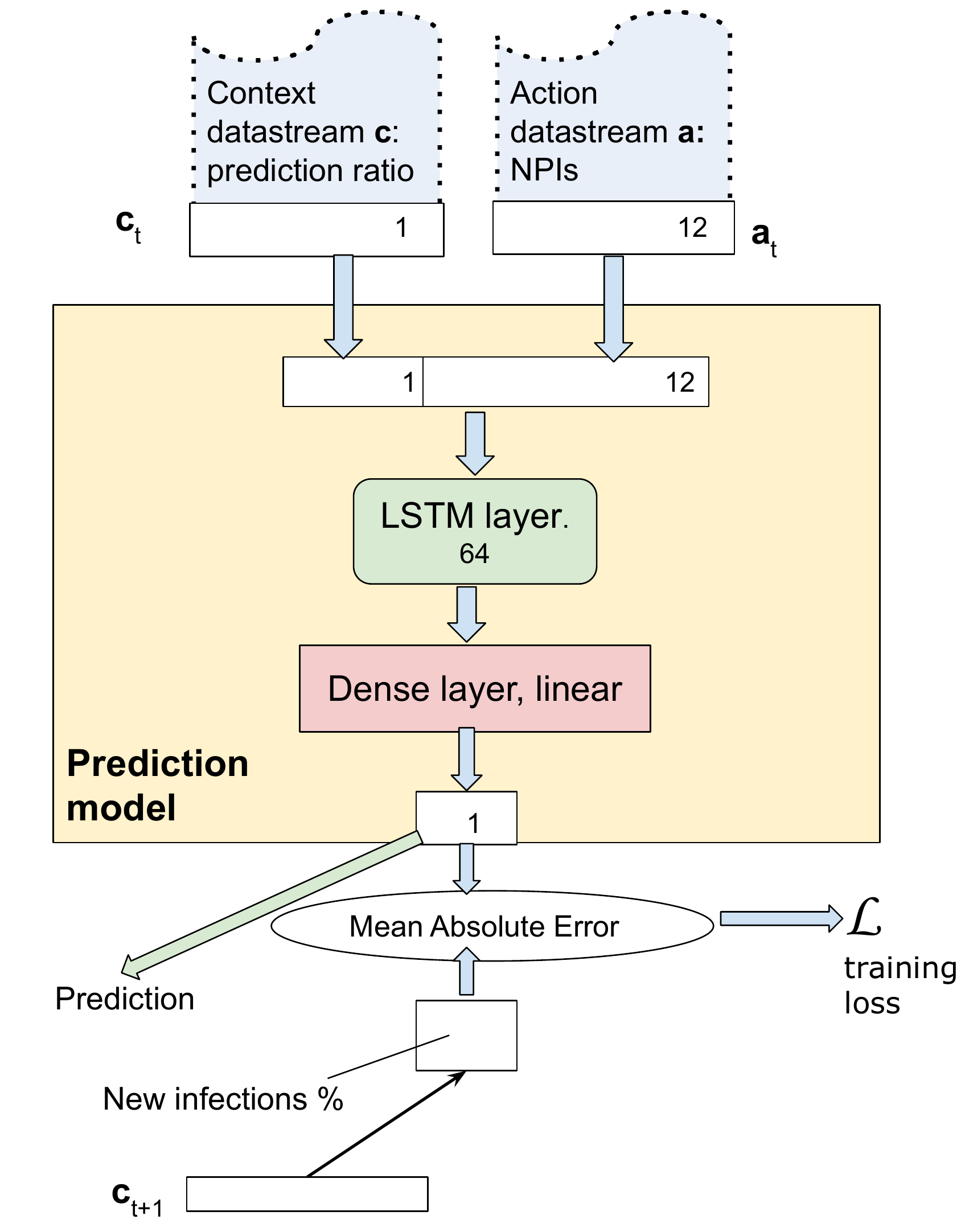}
    \vspace{2mm}
    \includegraphics[width=0.4\textwidth]{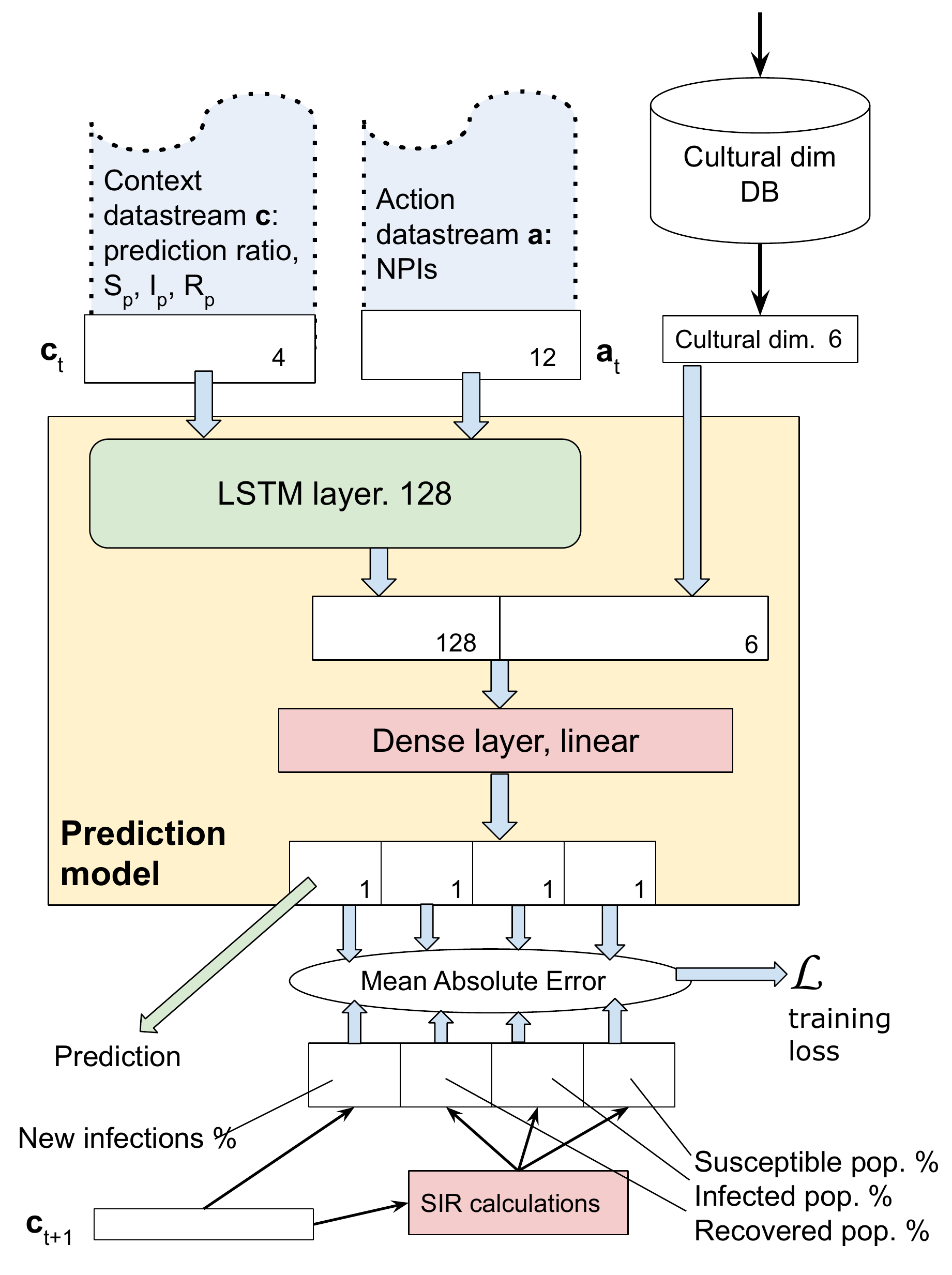}
    \includegraphics[width=0.4\textwidth]{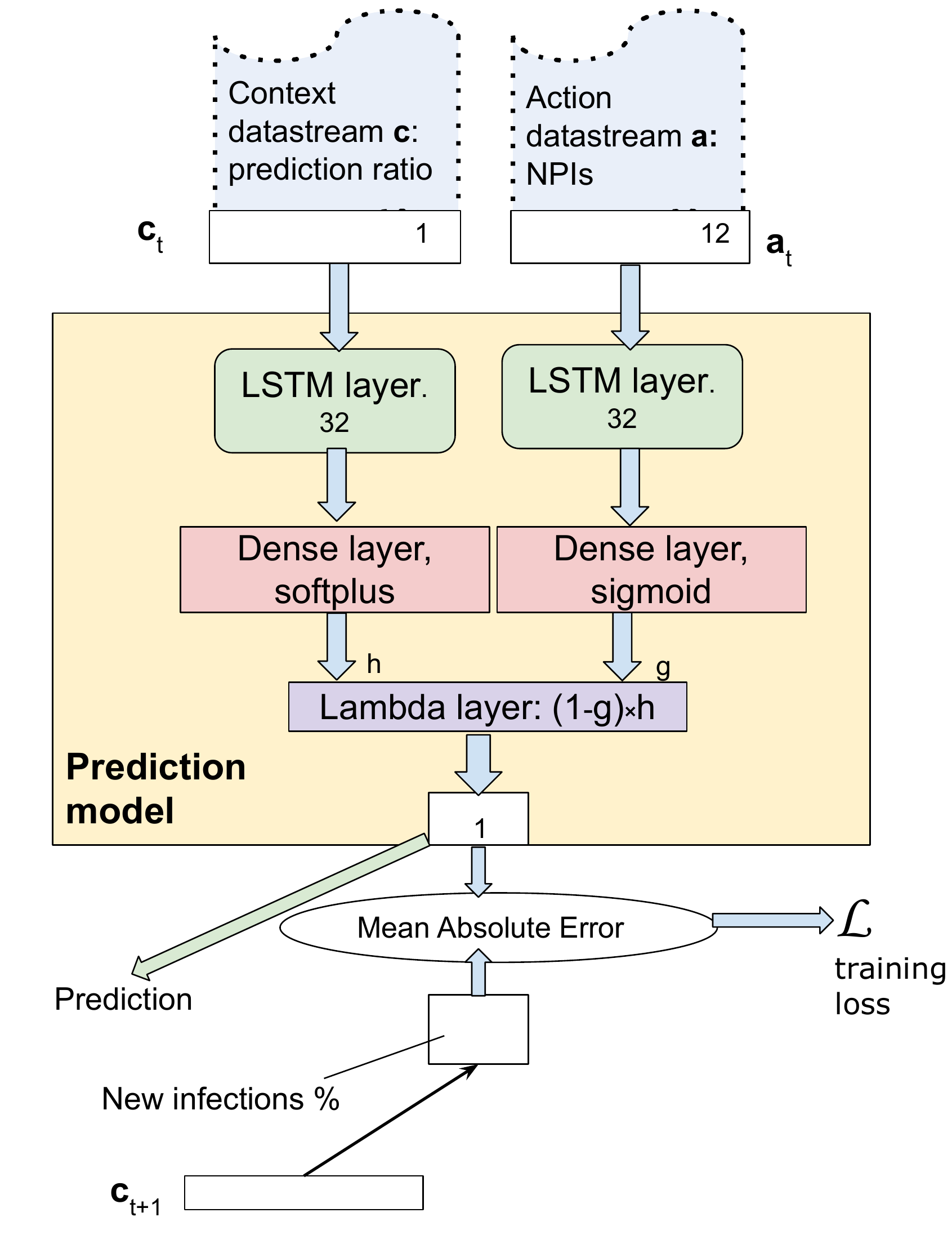}
    \includegraphics[width=0.4\textwidth]{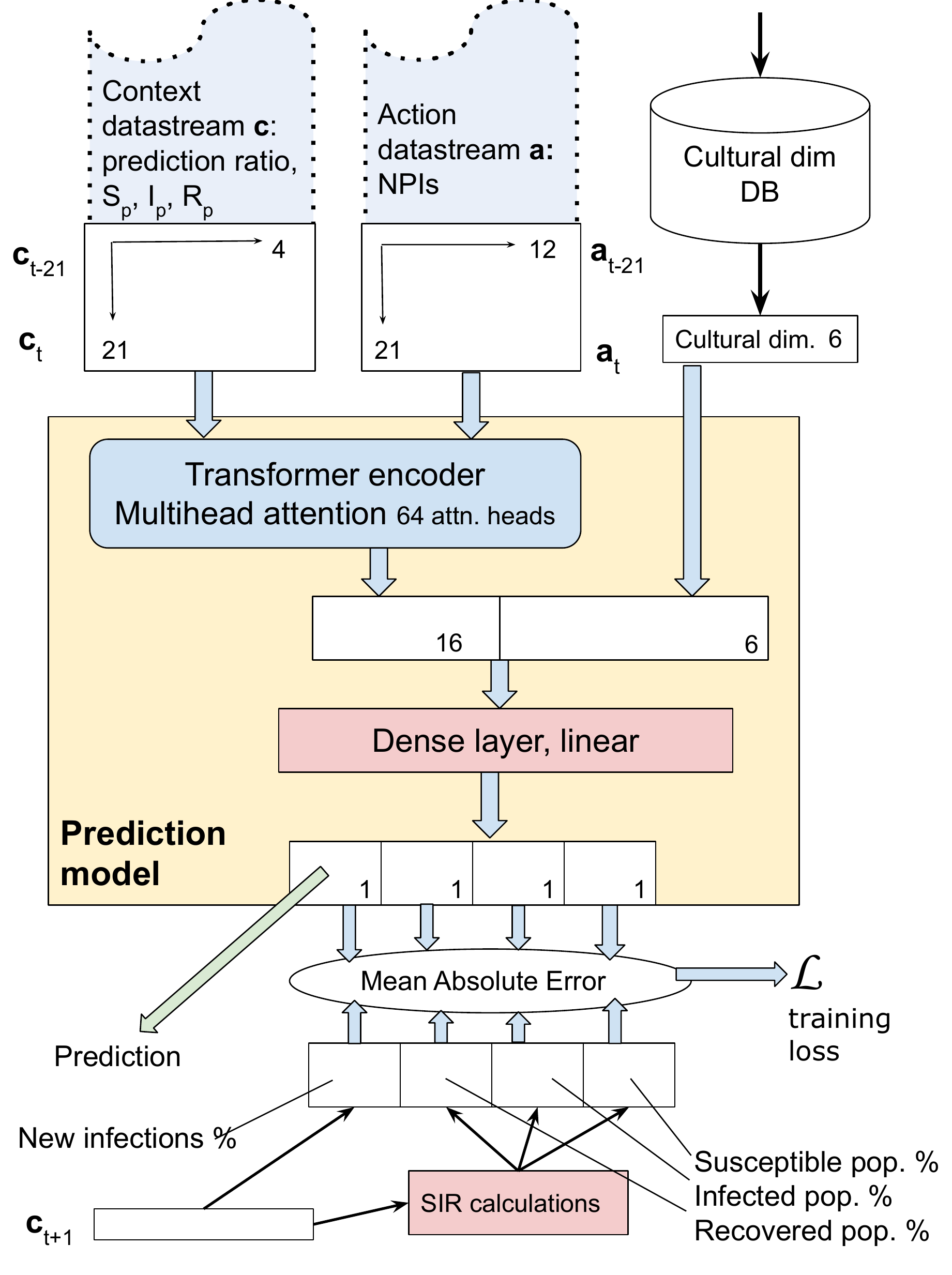}
    \caption{The architecture of the compared models: LSTM-Baseline (top-left), LSTM-UT-Cogn (bottom-left), LSTM-CultD-SIR (top-right) TRANSENC-CultD-SIR (bottom-right).}
    \label{fig:preds_methods}
\end{figure*}

The model, shown in Figure~\ref{fig:preds_methods}-bottom-left, is unusual in that it is using two separate branches for the context and the action data, with the predicted value being the proportion of new people infected from the population that is currently not infected (naturally, the absolute number of infected people can be calculated from this value). The input of the first branch is the infection ratio from the context stream which is processed by an LSTM layer, followed by a dense layer with one node and soft plus activation. In the second branch, the model takes as inputs the NPIs from the action stream, which is processed by a LSTM followed by a dense layer with a single node and sigmoid activation function. The outputs of context branch $h$ and action branch $g$ are combined using a lambda layer implementing the formula $(1-g) \times h$ to produce the output of the model. The model was trained on sequences of length of 21 days. 

\bigskip

The remaining three models were developed by our team, partially as part of our competition entry, partially through later improvements. 


\subsubsection{LSTM-Baseline}

The simplest, baseline model we are proposing also uses a LSTM network which in recent years had became the most popular way in the deep learning community to process data from time series that is presented to the model one at a time. In this, simplest model (Figure~\ref{fig:preds_methods}-top-left), we investigate the hypothesis that the LSTM network can learn how to select the important information from the combined context and action stream without any further input from the modeler. Before inputting our data to LSTM, we have to preprocess it such that the values are normalized. To achieve this, we use the same ``Infection Ratio'' column that is evaluated as follows: First, we compute the infected proportion by dividing the number of cases by population in region $r$ and day $t$. In other words, we compute $\mathit{a^{r}_{t}} = \frac{\numcases^{r}_{t}}{\population^{r}_{t}}$. Then, the smoothed version of $\mathit{a^{r}_{t}}$ is computed for each day by getting the average of these values in a 7-day time window. 
Next, we compute the percent change by $\predictionratio^{r}_{t + 1} = \frac{a^{r}_{t + 1} - a^{r}_{t}}{a^{r}_{t}} $


The model processes streams of data of a width of 13, with the first column being the value for the infection ratio while the rest of the columns representing the NPIs.
This input goes into an LSTM node with 64 nodes, and they are processed by a Dense layer with just one neuron that outputs ``Infection Ratio'' for the next day. We use the L1 loss between the output of the model and the real value of the ``Infection Ratio'' to train the network. During the inference, we only have access to NPI columns on each day in the future and use the prediction of the network as the ``Infection Ratio'' input for the following days. We also clip the network's output between 0 and 2 during inference to make sure that the outputs do not diverge. This is especially important when we use the model for longer predictions.

\subsubsection{Taking into account culture}

It had been an important part of the public narrative of the pandemic that various aspects of the interventions such as mask wearing, refraining from large gatherings, adherence to social distancing rules and vaccinations are culture-dependent. 

Unfortunately, quantifying various aspects of the culture as relates to the pandemic is not easy. Furthermore, similar cultures can accommodate very different public policies, as illustrated by the case of Scandinavian countries where culturally similar countries like Sweden and Norway chose to adapt different policy approaches. 
Nevertheless, the hypothesis that taking the cultural aspects into consideration can improve the prediction accuracy is definitely worth considering. A problem in implementing such a system is that in the social sciences culture is often discussed in qualitative, narrative form. There are relatively few examples of quantitative models of human interactions. One of the efforts that had attempted to assign numerical values to aspects of the culture of various nations is that cultural dimensions model~\cite{hofstede1984culture} which attempted to quantify natural culture along six numerical dimensions. Public databases are available at a nation-state level. While this model had received significant criticism over the years, among other things for the choice of a nation as the resolution of the model. For instance, the model does not differentiate between California and Alabama in the United States. However, to our best knowledge this is the only culture quantification model for which public databases exist for the majority of regions. 

We note that we do not make any assumptions about the impact of the cultural dimension values on our prediction - we add these values to the system and allow it to learn their possible impact. 

\subsubsection{Adding compartmental models}
We propose a method that leverages compartmental models that are mathematical models used for predicting pandemics.
 The proposed method not only does allow us to combine data available with mathematical models, but also generalizes to evolving nature of the pandemic since the trend is learned through available data. This allows us to better address the impact of decisions such as school closing. Furthermore, this approach is preferred when there is not much data since instead of learning the whole pattern in a black box model, we try to estimate the parameters of a compartmental model which is well studied and can describe why the model believes the trend is going to change in a particular way. In our implementation, we consider SIR models. SIR model assumes that the number of people are fixed. We denote the population by $P$. Everyone is in one of three states: Susceptible (S), Infected (I), and Recovered (R). 
 We add these columns to the data for each country using the following equations:
 \begin{equation}
 \label{eq:S_from_data}
     S_t = S_{t-1} - \textit{newCases}_t
 \end{equation}
 \noindent where $S_{0} = \textit{Population}$
 \begin{equation}
     I_t = I_{t-1} - \frac{1}{d}  \times I_{t-1} + \textit{newCases}_t - \textit{newDeaths}_t
 \end{equation}
\noindent where $1 / d$ is the daily recovery rate and $d$ is the average number of days required to recovering.
 \begin{equation}
     R_{t} = \textit{Population} - S_{t} - I_{t} 
 \end{equation}
 The transition from these states can be modeled by parameters $\alpha$ and $\beta$ as described below:
\begin{align}
S' &= -\alpha \times S \times I \\
I' &= \alpha \times S \times I - \beta \times I \\
R' &= \beta \times I,
\end{align}

\noindent where $S'$, $I'$, and $R'$ are the rate of change in value of $S$, $I$, and $R$ respectively. 
We train both LSTM and Transformer networks to take input from last $T$ days and predict the value of $S_p$, $I_p$, $R_p$ which are susceptible fraction of the population, infected fraction of the population and recovered fraction of the population for the next day(s). The model looks at NPI and all other data and outputs $S_p$, $I_p$, and $R_p$.

\subsubsection{LSTM based predictor using cultural dimensions and the SIR model}
In this model, we use compartmental model to create new columns susceptible fraction of the population ($S_p$), infected fraction of the population ($I_p$), and recovered fraction of the population ($R_p$). We initialize $S_p$ with 1, and $I_p$ and $R_p$ with 0. Then, we calculate these values for the next rows based on equations 1, 2, and 3 over population for each country or region. We use 14 as the average number of days required for recovery to compute recovery rate in equation 2.
We use these columns alongside the infection ratio column as context input. We concatenate the context input and action input (NPI columns) of 21 previous days  and feed it to a LSTM layer. Then, we concatenate new features such as cultural features of Hofstede dimensions as constant features to the output of the LSTM layer and feed them to a dense layer with 4 nodes. The model is trained with Adam optimizer and mean absolute error and outputs the infection ratio, $S_p$, $I_p$, and $R_p$.
See Figure~\ref{fig:preds_methods}-top-right.

\subsubsection{Transformer encoder based predictor using cultural dimensions and the SIR model}
This model is similar to LSTM-CultD-SIR, but we use a transformer encoder layer instead of the LSTM layer. The difference is that this model can read the whole sequence all at the same while the LSTM-CultD-SIR model reads that sequence one by one. 
Transformers were first introduced in~\cite{vaswani2017attention}. Transformer is a model architecture that instead of using recurrent networks uses an attention mechanism to learn relations between input and output. The main advantage of the transformers is their ability to see the sequence of data in parallel and learn very long-term interactions. We propose using the multi-head attention module from the transformer architecture to train the predictor model. To the best of our knowledge, we are the first to use attention models on NPI features and combining the attention model's output with the cultural dimensions and the SIR model for prediction of new Covid-19 cases. 

The transformer layer includes attention and normalization part and a feed forward part. The attention and normalization part includes a multi head attention layer, dropout and normalization layer. The feed forward layer is a sequence of two dense layers one with ReLU activation and the other one with linear activation, a dropout layer and a normalization layer.
See Figure~\ref{fig:preds_methods}-bottom-right.

\section{Experimental Studies}
\label{experiments}

To evaluate the predictive accuracy of the model, we need a metric that smooths out daily variations and is comparable across regions with different populations. Thus for a region $r$ with a population $P_{r}$ we will use the average number of cases per 100k people over a span of 7 days:
\begin{equation}
    \textrm{\small{Cumul-7DMA-MAE-per-100K}}_{(r)} = \sum\limits_{\tiny{d \in \textrm{D}}}^{}{|\bar{y} - \bar{\hat{y}}|} \times \frac{100000}{P_{r}},
\end{equation}

\noindent where $\bar{x}$ is the 7-day moving average on $x$ and  denotes the population in region $r$.

\begin{figure}
    \centering
    \includegraphics[width=0.85\columnwidth]{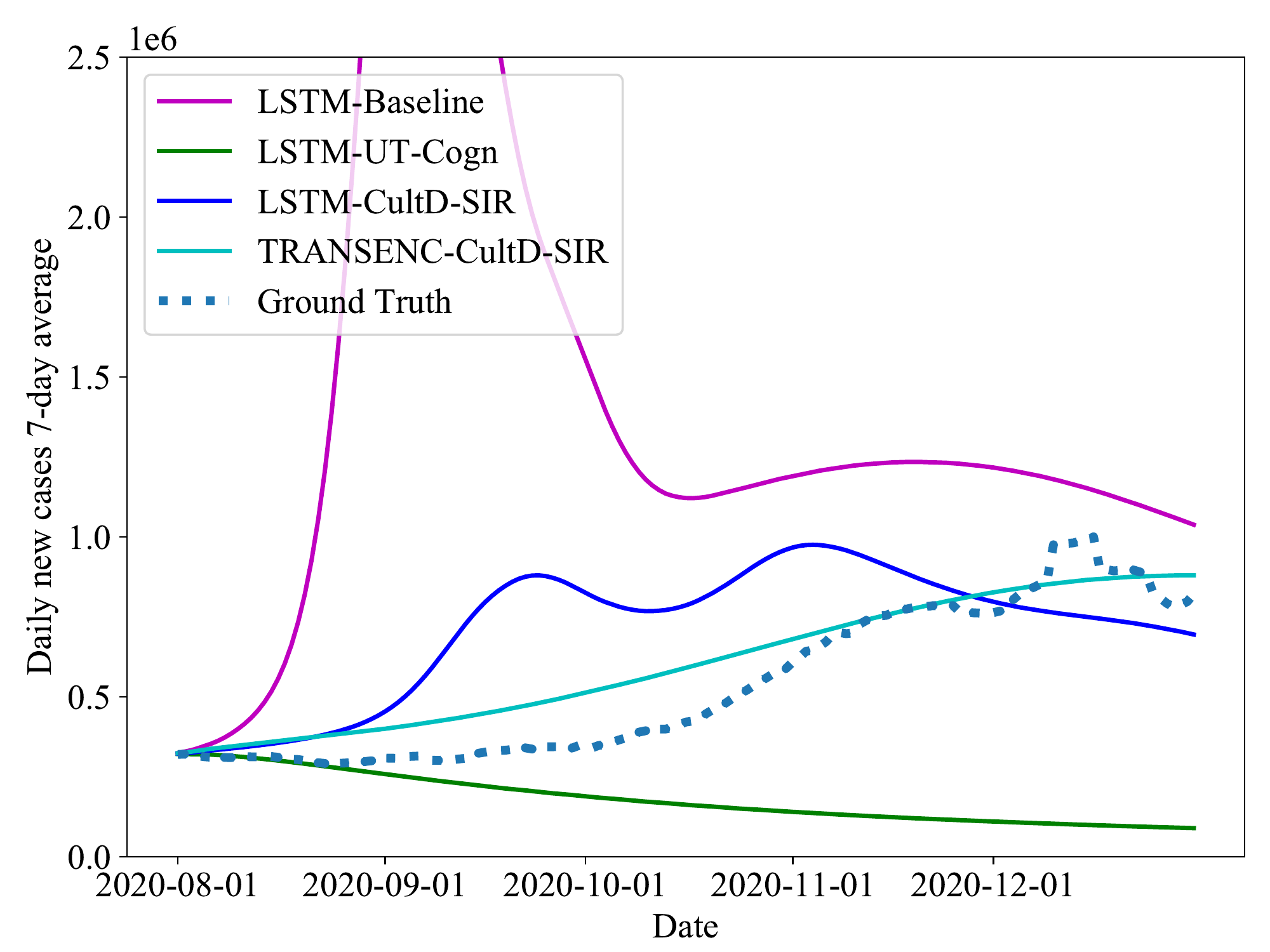}\\
    \includegraphics[width=0.85\columnwidth]{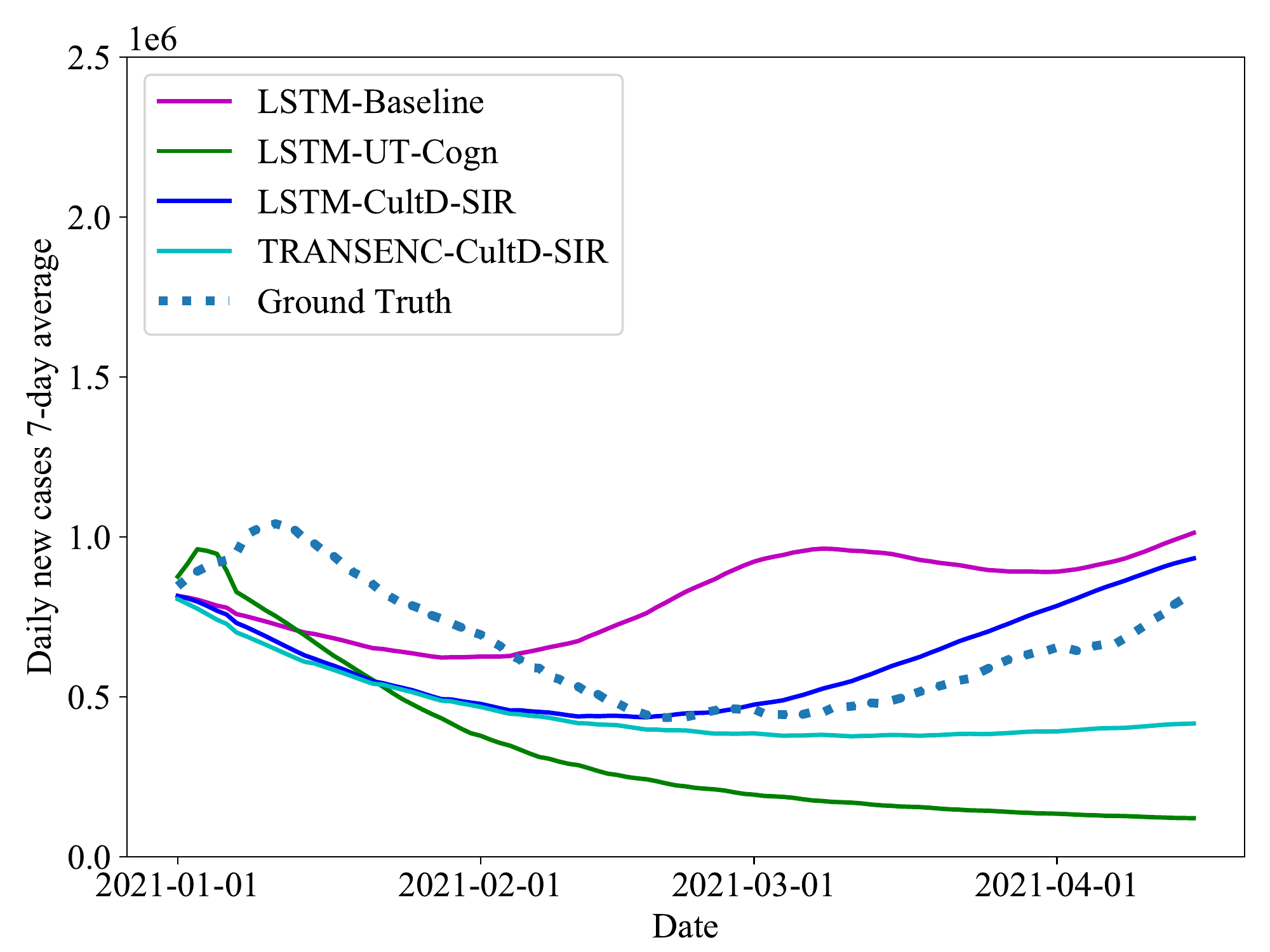}

    \caption{Average of 7-day predicted daily new cases over all countries using our predictors, LSTM-CultD-SIR and TRANSENC-CultD-SIR and two baselines LSTM-Baseline and LSTM-UT-Cogn. \textbf{Top:} E2020,    \textbf{Bottom:} E2021.}
    \label{fig:preds}
\end{figure}

\begin{figure}
    \centering
    \includegraphics[width=0.85\columnwidth]{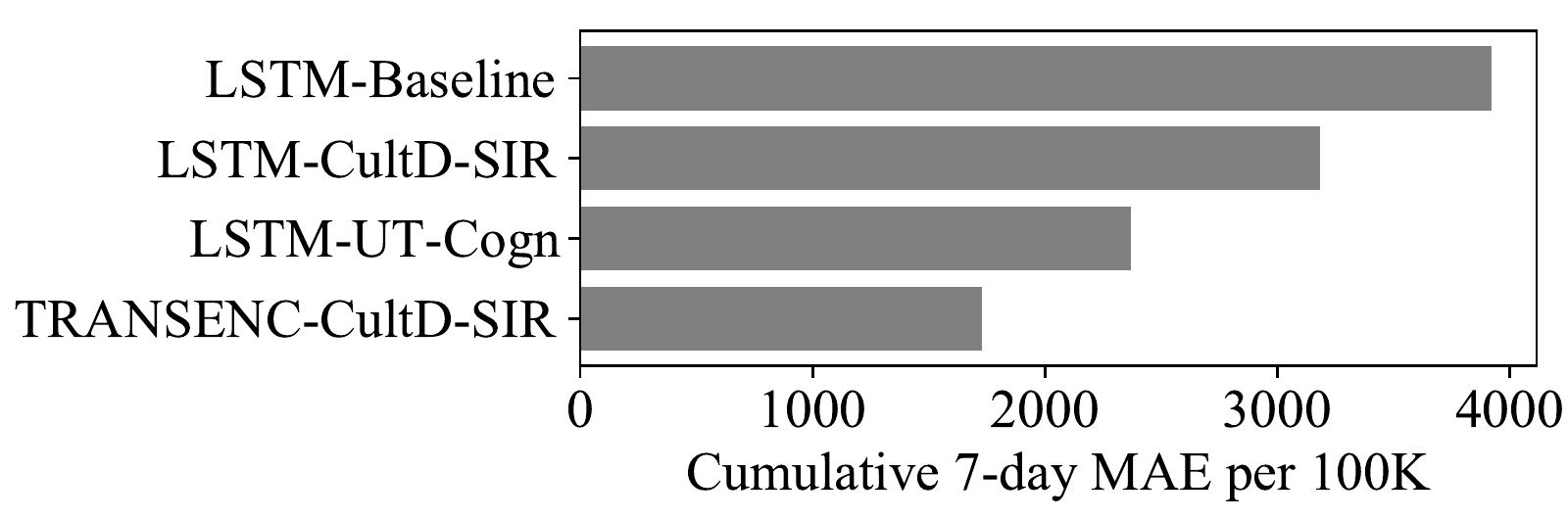}
    \includegraphics[width=0.85\columnwidth]{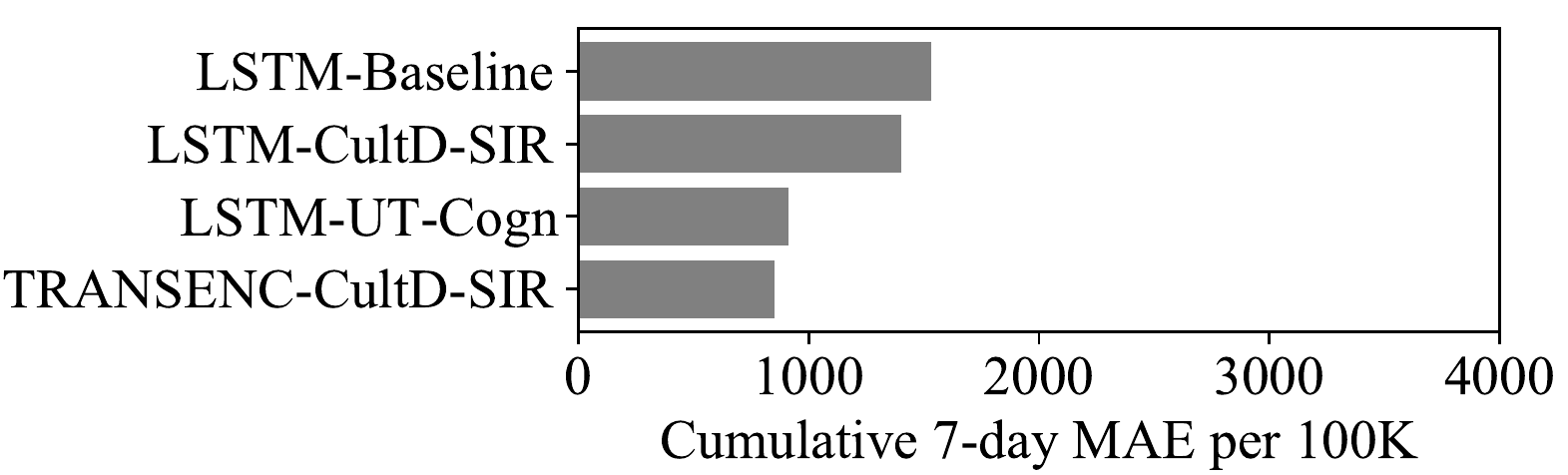}
    \caption{Cumulative 7-day mean absolute error per 100k for each prediction approach.
    \textbf{Top:} E2020. \textbf{Bottom:} E2021.}
    \label{fig:mae}
\end{figure}

\begin{figure}
    \centering
    \includegraphics[width=0.85\columnwidth]{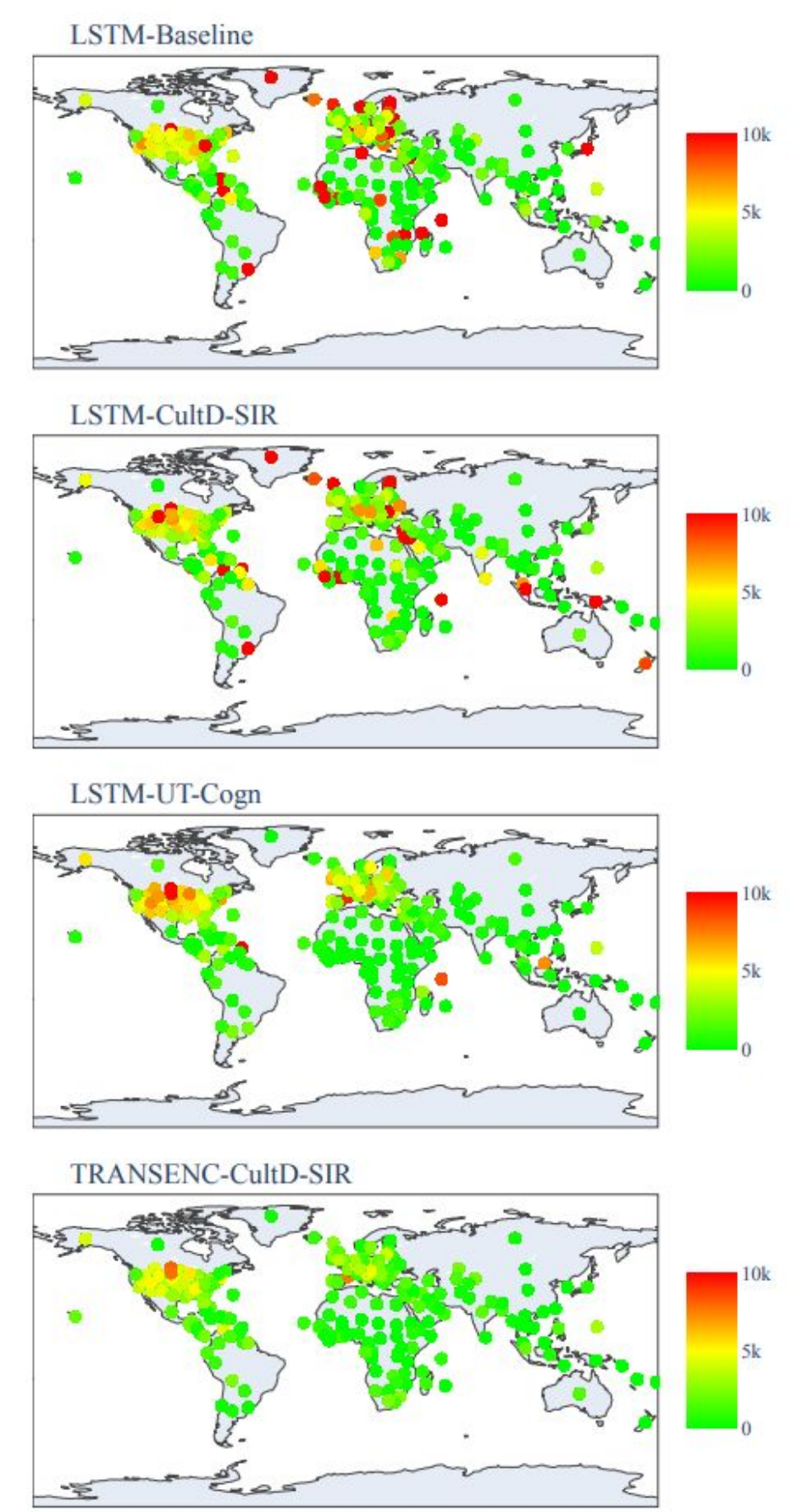}
    \caption{Color scaled cumulative 7-day mean absolute error per 100k per country or region based on each prediction approach. Green color shows zero to 2k and red color shows 8k or more cumulative 7-day mean absolute error per 100k.}
    \label{fig:mae-global}
\end{figure}

We trained all the models separately by using a few months of data from all the countries and regions. 
For every model, we use 1000 epochs for training with early stopping with patience 20 that restores best weights. We split the training data into training and validation with 90\% and 10\% rate, respectively, with a batch size of 32. The learning rate is also 0.001. 

We run two different experiments, with the training data and test data being chosen from different calendar months. Experiment E2020 used data from January 2020 to July 2020 for training and data from August 2020 to end of December 2020 for evaluation. Experiment E2021 used data from the whole year of 2020 and made predictions for January to April 2021. See Figure~\ref{fig:preds}.

Based on our experiments, our TRANSENC-CultD-SIR approach had the lowest cumulative mean absolute error per 100k over 7 days. See Figure~\ref{fig:mae} for the overall cumulative 7 day moving average mean absolute error per 100k for both of the experiments. Also, Figure~\ref{fig:mae-global} shows the color scaled version of this metric for all the countries and regions around the world for E2020 experiment. The green nodes are showing the regions or countries with the lowest error (0 to 2k) per 100k, and red nodes are showing 8k or more error per 100k. We find that the TRANSENC-CultD-SIR approach has the lowest number of red-orange nodes which means that it has the better performance comparing to other approaches.

\section{Conclusion}
\label{conclusion}

In this paper, we described the design of several pandemic prediction models and compared them with each other. As a comparison point, we used the model that was used as the official predictor for the finals of the XPrize Pandemic Response Challenge. The models introduced in this paper build on and improve our submission to this competition. By testing the models on data that extends several months {\em after} after the competition, we can make several observations that can serve as lessons for modeling approaches in the future. First, models that are finely tuned to predict over the spans of days and weeks accurately can diverge significantly over the span of months. Second, sophisticated machine learning models such as transformer-style multi-head attention replacing LSTMs can produce an iterative improvement if everything else is equal but are not making a decisive difference in prediction accuracy. Third, while the canonical models of prediction such as the SIR compartmental model cannot, by themselves, provide an accurate prediction, they can serve a useful role in preventing runaway errors in the models. Finally, while cultural factors are clearly influencing the evolution of the pandemic, we do not yet have a method to incorporate this information in a rigorous manner. 
\bibliography{IEEEabrv,References}
%



\end{document}